\PassOptionsToPackage{square, comma, sort&compress}{natbib}

\documentclass[review]{elsarticle}

\usepackage{amsmath}
\usepackage[colorlinks=true,linkcolor=blue,urlcolor=blue,citecolor=blue]{hyperref}
\usepackage{graphicx}
\usepackage[squaren, Gray, cdot]{SIunits}
\usepackage{amssymb}

\usepackage{tikz}

\renewcommand*\textcircled[1]{\tikz[baseline=(char.base)]{
            \node[shape=circle,draw,inner sep=0.5pt] (char) {#1};}}

\begin{document}

\begin{frontmatter}

\title{Circular dichroism in atomic vapors: magnetically induced transitions responsible for two distinct behaviors}

\author[IPR]{Armen Sargsyan}
\author[IPR]{Arevik Amiryan}
\author[IPR,NCU]{Ara Tonoyan}
\author[JGU]{Emmanuel Klinger\corref{mycorrespondingauthor}}
\cortext[mycorrespondingauthor]{Corresponding author}
\ead{eklinger@uni-mainz.de}
\author[IPR]{David Sarkisyan}

\address[IPR]{Institute for Physical Research – National Academy of Sciences of Armenia, 0203 Ashtarak-2, Armenia}
\address[NCU]{Institute of Physics, Faculty of Physics, Astronomy and Informatics, Nicolaus Copernicus University, Grudzi\c{a}dzka 5, PL-87-100 Toru\'n, Poland}
\address[JGU]{Helmholtz-Institut Mainz -- GSI Helmholtzzentrum f{\"u}r Schwerionenforschung, Johannes Gutenberg-Universit{\"a}t, D-55128 Mainz, Germany}

\begin{abstract}
Atomic transitions of alkali metals for which the condition $F_e-F_g = \pm2$ is satisfied have null probability in a zero magnetic field, while a giant increase can occur when an external field is applied. Such transitions, often referred to as magnetically-induced (MI) transitions, have received interest because their high probabilities in wide ranges of external magnetic fields which, in some cases, are even higher than that of usual atomic transitions.  Previously, the following rule was established: the intensities of MI transitions with $\Delta F=\pm2$ are maximum when using respectively $\sigma^\pm$ radiation. Within the same ground state, the difference in intensity for $\sigma^+$ and $\sigma^-$  radiations can be significant, leading to magnetically induced circular dichroism (MCD), referred to as type-1. Here, we show that even among the strongest MI transitions, $i.e$ originating from different ground states for $\sigma^+$ and $\sigma^-$, the probability of MI transition with $\Delta F = + 2$ is always greater, which leads to another type of MCD. Our experiments are performed with a Cs-filled nanocell, where the laser is tuned around the D$_2$ line; similar results are expected with other alkali metals. Theoretical calculations are in excellent agreement with the experimental measurements.
\end{abstract}


\end{frontmatter}


\section{Introduction}
Interest in magnetically-induced transitions of alkali metal atoms is caused by the high probabilities, in wide ranges of external magnetic fields, that these transitions can reach, as well as by their large frequency shifts with respect to unperturbed atomic transitions \cite{tremblayPRA1990,sargsyanLPL2014,scottoPRA2015,sargsyanJETPL2017,tonoyanEPL2018,sargsyanOS2019,sargsyanJPB2020}. According to the selection rules for transitions between the ground and excited hyperfine levels  (in the dipole approximation), one must have $\Delta F =F_e - F_g= 0, \pm 1$. The probabilities of the transitions for which the condition $\Delta F = \pm2$ is satisfied  are null at zero magnetic field, while a giant increase of their probabilities can occur in an external magnetic field. Such transitions of alkali metal atoms are referred to as magnetically induced (MI) transitions \cite{sargsyanJETPL2017,tonoyanEPL2018,sargsyanOS2019,sargsyanJPB2020}. A striking example of a giant increase in probability, in particular, is the behavior of transitions $F_g = 3\rightarrow F_e = 5$ (seven transitions) of Cs D$_2$ line, transitions $F_g = 2\rightarrow F_e = 4$ (five transitions) of $^{85}$Rb D$_2$ line or  transitions $F_g=1\rightarrow F_e=3$ (three transitions) of $^{87}$Rb D$_2$ line in an external magnetic field. The modification of the probabilities of the atomic transitions is caused by the ``mixing'' effect of the magnetic sublevels induced by the presence of a magnetic field \cite{tremblayPRA1990,auzinshBook2010}.  We have previously established the following rule: the intensities of MI transitions with $\Delta F = \pm 2$ are maximum when using respectively $\sigma^\pm$ radiation \cite{sargsyanJETPL2017,tonoyanEPL2018}. For some MI transitions, the difference in intensity when using $\sigma^+$ or $\sigma^-$ radiation can reach several orders of magnitude, that is an anomalous magnetically-induced circular dichroism (MCD) occurs, which we refer to as type-1. 

Magneto-optical processes are widely used, their applications include magneto-optical tomography, narrowband atomic filtering, optical magnetometry, tunable laser frequency locking, etc. \cite{budkerRMO2002,zentileJPhysB2014,keaveneyOL2018,taoOE2019,sargsyanOL2014,mathewOL2018,bhattaraiPLA2019,bhushanPLA2019,klingerAO2020}. These experiments often make use of circular dichroism in alkali vapors. It is therefore necessary to get a thorough understanding of MCD, which could enhance a wide amount of these applications. Let us note that a strong MCD has been measured for the number of electrons ejected by multiphoton ionization of resonantly excited He$^{+}$ ions using circulary-polarized light \cite{ilchenPRL2017}.
 In the present work, we reveal another type of MCD: even among the strongest MI transitions, $i.e$ originating from different ground states for $\sigma^+$ and $\sigma^-$, the probability of MI transition with $\Delta F = + 2$ is always greater than that of MI transition with $\Delta F = - 2$. For Cs atoms, the amplitude of the MI transition $|3,-3\rangle \rightarrow |5,-2\rangle$ of the D$_2$ line (strongest for $\sigma^+$ radiation) is shown to be larger than $| 4,-1\rangle\rightarrow |2,-2\rangle$ (strongest for $\sigma^-$ radiation), in the range 0.2 to 5~kG. Because this dichroism occurs between transitions originating from different ground states, we refer to it as type-2 MCD.

To observe this unusual behavior of magnetically induced optical transitions, we use a Cs-filled optical nanocell (NC) illuminated with a laser tuned in the vicinity of the D$_2$ line. Sub-Doppler resolution with a simple single beam geometry, providing a linear response in atomic media for absorption experiments, was previously demonstrated using NC with a thickness $\ell$ of the order of the resonant radiation half-wavelength \cite{sargsyanJPB2016}. The second derivative (SD) method is used to further narrow the spectral width of atomic transitions \cite{sargsyanOL2019} since high spectral resolution is necessary to distinguish closely frequency spaced transitions in intermediate magnetic fields. The article is organized as fallows: in Sec.~\ref{sec:2}, we summarize the theoretical model used to calculate transitions frequencies and intensities, and spectral lineshapes. In Sec.~\ref{sec:3}, we detail the experimental setup, and finally we discuss the results in Sec.~\ref{sec:4}.

\section{Theoretical model}
\label{sec:2}
\subsection{Alkali atoms in a magnetic field}
\label{sec:TheoryAlkaliMagneticField}

It is well known that, due to the presence of a static magnetic field, atoms have their energy levels split, and their optical transitions undergo shifts and change of amplitude. As the magnetic field couples hyperfine states according to the selection rules $\Delta L=0$, $\Delta F=\pm 1$, $\Delta m_F=0$ \cite{tremblayPRA1990}, the Hamiltonian matrix describing the evolution of the hyperfine levels is not diagonal and to find the perturbed energy levels, one has to diagonalize it. As a results of the diagonalization, the shifted energy levels are the diagonal elements, and the state vectors read
\begin{equation}
|\psi(F'_g,m_{F_g})\rangle = \sum _{F_g}C_{F'_{g}F_{g}}|F_{g},m_{F_{g}}\rangle,
\end{equation}
where the label $g$ refers to ground states, a similar relation for excited states can be written by swapping $g$ for $e$; $C_{F'F}$ are the (magnetic field dependent) mixing coefficients. The intensity $A_j$ of a transition, labelled $j$, between the states $|F_g',m_{F_g}\rangle$ and $|F_e',m_{F_e}\rangle$ is proportional to $a^2[\psi(F_e',m_{F_e});\psi(F_g',m_{F_g});q]$, where
\begin{equation}
a[\psi(F_e',m_{F_e});\psi(F_g',m_{F_g});q] =\sum_
{F_e,F_g}C_{F'_eF_e}a(F_e',m_{F_e};F_g',m_{F_g};q)C_{F'_gF_g},
\end{equation}
and $q=0,\pm1$ is associated with the polarization of the incident electric field. The coefficient $a(F_e',m_{F_e};F_g',m_{F_g};q)$ reads 
\begin{equation}
\begin{aligned}
a(F_e',m_{F_e};F_g',m_{F_g};q)=(-1)^{1+I+J_e+F_e+F_g-m_{Fe}}\sqrt{2J_e+1}\sqrt{2F_e+1}\\ \times \sqrt{2F_g+1} \left( \begin{array}{r@{\quad}cr} 
F_e & 1 & F_g \\
-m_{F_e} & q & m_{F_g}
\end{array}\right)
\left\{ \begin{array}{r@{\quad}cr} 
F_e & 1 & F_g \\
J_g & I & J_e
\end{array}\right\},
\end{aligned}
\end{equation}
where the parentheses and the curly brackets denote the 3-$j$ and 6-$j$ coefficients, respectively.

\begin{figure}[h!]
\centering
\includegraphics[scale=1]{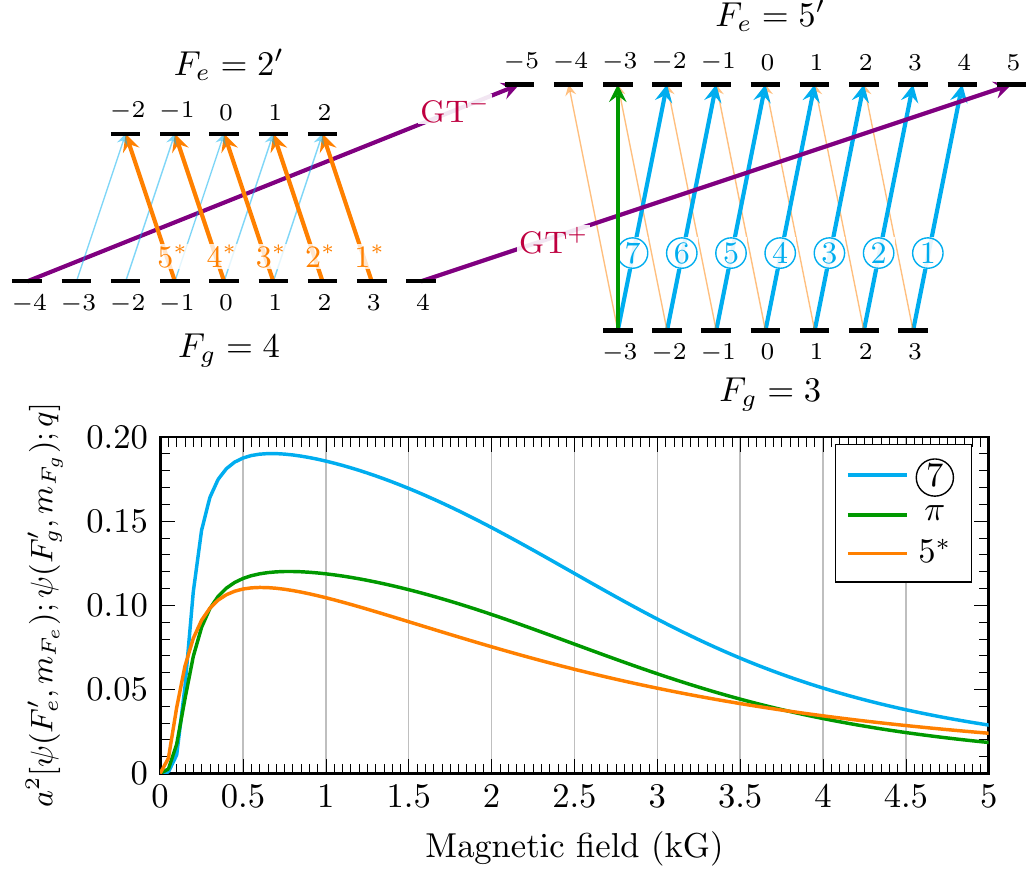}
\caption{Calculated probabilities of the strongest MI transitions of Cs D$_2$ line: (cyan) $\sigma^+$ polarization, (orange) $\sigma^-$ polarization and (green) $\pi$ polarization. The inset shows the transition diagrams and labels, where the states $|3',m_F'\rangle$ and $|4',m_F'\rangle$ have been omitted; blue lines are $\sigma^+$ MI transitions, orange lines are $\sigma^-$ MI transitions, thin lines depict weak MI transitions. Only the strongest MI transition from $\pi$ polarization was drawn (green line). Purple lines show the guiding transitions, their amplitude is always 0.25.}
\label{fig:1}
\end{figure}

Figure~\ref{fig:1} shows the calculated intensities of the strongest MI transitions for each polarization:  $|3,–3\rangle \rightarrow |5',–2'\rangle$ labelled $\textcircled{7}$ ($\sigma^+$ radiation, $q=+1$); $|4,–1\rangle \rightarrow |2',–2'\rangle$  labelled $5^*$ ($\sigma^-$ radiation; $q=-1$); and $|3,–3\rangle \rightarrow |5',–3'\rangle$  ($\pi$ radiation, $q=0$). As is seen, the probability of transition $\textcircled{7}$ is the strongest among other MI transitions. 

\subsection{Absorption profile}
Previous works have shown that the atomic response from the NC is an interferometric combination of bulk reflection and transmission.  For a dilute vapor confined in  a cell whose length verifies $\ell \ll  u/(2\pi\Gamma)$ where $u(\Theta)=\sqrt{2k_B\Theta/m_a}$ is the thermal velocity, the resonant contribution to the weak-probe transmission profile can be approximated by
\begin{equation}
S_t(\nu,\ell,B)=\frac{2t(1-r)(1-r^2)E^2}{|Q|^2}\cdot \text{Im}\Big[\sum_j\chi_j(\nu, \ell, B)\Big],
\end{equation}
with $Q=1-r^2\exp[2ik\ell]$, $r$ and $t$ are field reflection and transmission coefficient respectively,  $E$ the probe field amplitude. The contribution of each transition $j$ to the transmission spectrum is given by
\begin{equation}
\chi_j(\nu,\ell, B) = -4\big(1-r\exp[ik\ell]\big)^2~\frac{\sin^2[k\ell/2]}{Q} \cdot\frac{\mathcal N}{ku\sqrt{\pi}} \cdot\frac{iA_j}{\Gamma/2 -i\Delta_j},
\end{equation}
where $\mathcal N$ is the number density of the vapor, $k=2\pi/\lambda$ is the probe field wave-number, $\Delta_j=\nu-\nu_j$ is the detuning of the laser with respect to the frequency of the atomic transitions, and $A_j$ and $\nu_j$ are respectively the magnetic field dependent amplitude and frequency of each transition, see section~\ref{sec:TheoryAlkaliMagneticField}. The homogeneous (half-width) broadening $\Gamma$ includes various contributions such as the natural linewidth, atom-wall collisions, or laser linewidth, and is left as a free parameter in our simulations. The SD spectrum is then simply computed numerically with a second order derivative:
\begin{equation}
S''_t=\frac{\partial^2}{\partial \nu^2} S_t 
\end{equation}
Note that, because $S_t$ is background-free, one can simply express the resonant contribution to the absorption spectrum as $S_a(\nu) = -1\times S_t(\nu)$.

\section{Experimental details}
\label{sec:3}
\subsection{The nanometric-thin cell}

The design of the two Cs-filled NCs used in our experiments is similar to that of extremely thin cell described earlier \cite{keaveneyPRL2012}. The rectangular 20 mm $\times$ 30 mm, 2.5 mm-thick window wafers polished to  about 1~nm surface roughness are fabricated from commercial sapphire (Al$_2$O$_3$), which is chemically resistant to hot alkali vapors up to 1000$^\circ$C. The wafers are cut across the c-axis to minimize the birefringence. In order to exploit variable vapor column thickness, the cell is vertically wedged by placing a 1.5~$\mu m$ thick platinum spacer strip between the window sat the bottom side prior to gluing. A thermocouple is attached to the sapphire side arm at the boundary of metallic Cs to measure the temperature, which determines the vapor pressure. The temperature at this point was set to about 100$^\circ$C, while the windows temperature was kept some 20$^\circ$C higher to prevent condensation. This temperature regime corresponds to the Cs number density $\mathcal N \approx 10^{13}$~cm$^3$. The NC operated with a special oven with two optical outlets. The oven (with the NC fixed inside) was rigidly attached to a translation stage for smooth vertical movement to adjust the needed vapor column thickness without variation of thermal conditions. Note that all experimental results presented below have been obtained with Cs vapor column thickness $\ell \sim\lambda/2 \approx 426~$nm, for which a strong spectral narrowing, the Dicke narrowing \cite{sargsyanJPB2016}, occurs.

\begin{figure}[h!]
\centering
\includegraphics[width=9cm]{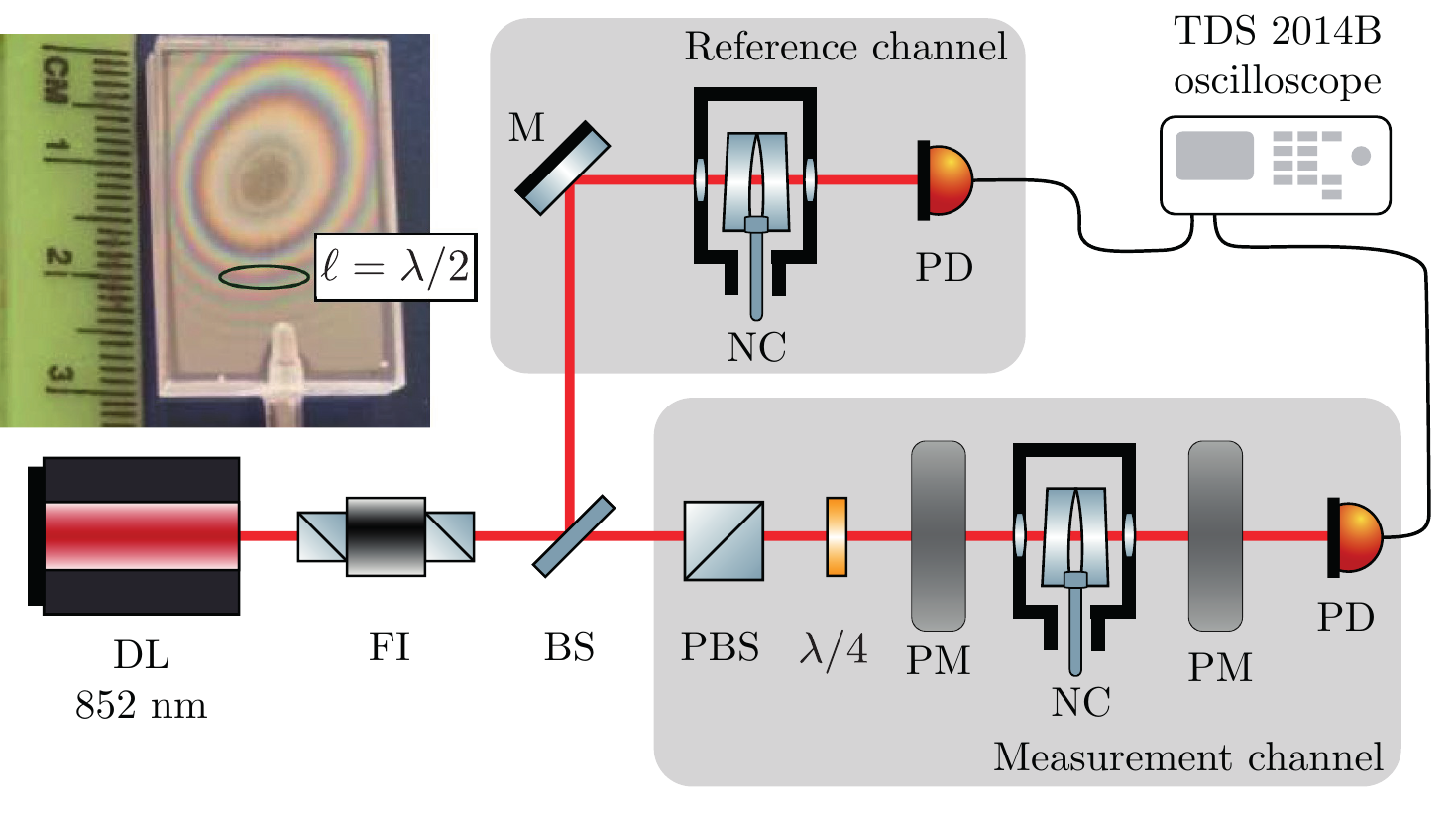}
\caption{Sketch of the experimental setup. DL –- tunable diode laser, FI –- Faraday isolator, $\lambda/4$ –- quarter-wave plate, PBS –- polarizing beam splitter, NC – nanocell in the oven; PM –-  permanent magnet, PD – photo-detectors. The reference channel is composed by an auxiliary Cs NC providing a frequency reference at zero magnetic field. The inset on the top left side show a picture of the Cs NC, where the region of thickness $\ell = \lambda/2$ has been highlighted.}
\label{fig:2}
\end{figure}

\subsection{Experimental setup}

Figure~\ref{fig:2} shows the layout of the experimental setup. A frequency-tunable cw diode laser with a wavelength $\lambda=852$~nm, protected by a Faraday isolator (FI), emits a linearly polarized radiation directed at normal incidence onto a Cs NC mounted inside the oven. A fraction of the incident light was sent by a beam splitter to a frequency reference channel composed by another Cs NC with no applied magnetic field. A quarter-wave plate, placed in between a polarizer and the NC, allows to switch between $\sigma^+$ (left-hand) and $\sigma^-$ (right-hand) circularly-polarized radiation. We have chosen to use a diode laser without external cavity for its large mode-hop free tuning range -- 20~GHz in contrast to only 5~GHz with an external cavity, which has however a larger linewidth $\gamma_l \sim  2\pi \cdot 10$~MHz. Even thought the laser additionally broadens the transitions, the SD method allows us to maintain the transition linewidth below 60~MHz. This choice was motivated by the fact that transitions originating from the same ground state span over more than 5~GHz for magnetic fields larger than 1~kG.

\section{Results and discussions}
\label{sec:4}

To show that dichroism occurs in the vapor because of $\sigma^+$ and $\sigma^-$ MI transitions having different intensities (see Fig.~\ref{fig:1}), we record  separately the absorption spectrum of $\sigma^+$ and $\sigma^-$ radiations at different magnetic fields. The SD method, which preserves both frequency positions and amplitudes of spectral features, is known to be a very convenient tool for quantitative spectroscopy and was used to improve our spectral resolution, allowing us to track the evolution of each transition individually.


\begin{figure}[h!]
\centering
\includegraphics[scale=1]{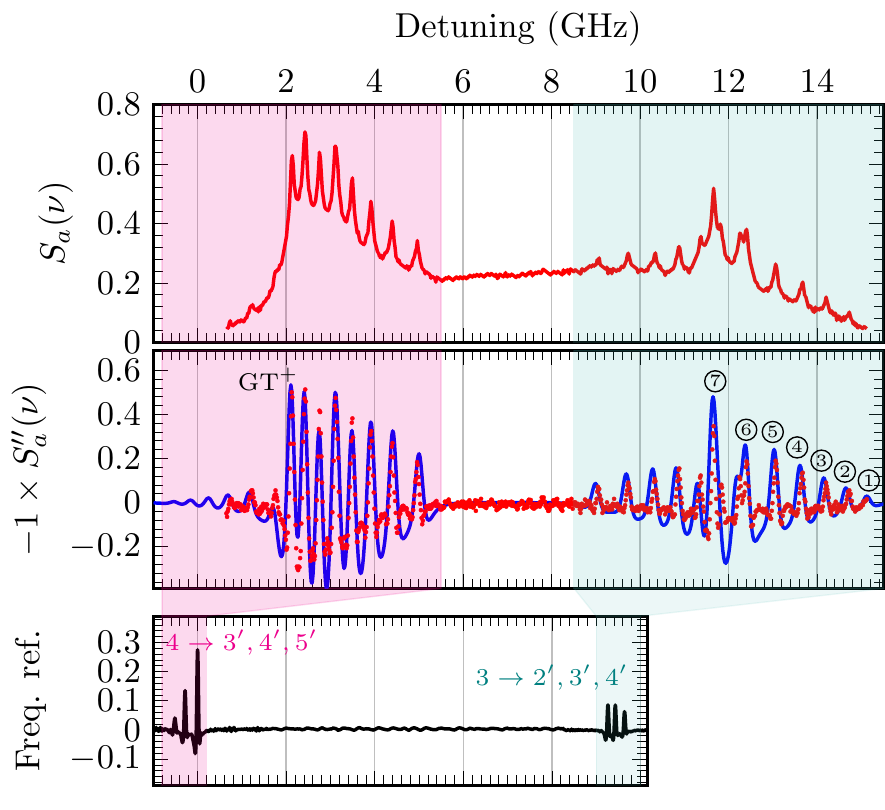}
\caption{Cs D$_2$ line spectrum of a $\sigma^+$ polarized radiation recorded at $B=1.50~$kG. Top panel: absorption spectrum; middle panel: SD of the absorption spectrum (red dots: experiment, blue solid line: theory); bottom panel: frequency reference.}
\label{fig:3}
\end{figure}

\subsection{MI transitions for a $\sigma^+$ excitation}

Figure~\ref{fig:3} shows the absorption spectrum (top panel) of the Cs D$_2$ line recorded from a NC with $\ell \sim\lambda / 2 \approx 426$~nm  and  illuminated with a left-circularly polarized laser radiation, at $B= 1.50$~kG. The middle panel shows the SD spectrum obtained after applying second order differentiation to the absorption spectrum, which shows a much improved spectral resolution. The blue solid line is the calculated spectrum corresponding to the experimental parameters, where the fitted linewidth  is about 40 MHz; note the good agreement between the theoretical and experimental spectra. The lowest curve (black solid line) is the frequency reference spectrum obtained from another NC with $\ell\sim\lambda/2\approx 426$~ nm at zero magnetic field.

\begin{figure}[h!]
\centering
\includegraphics[scale=1]{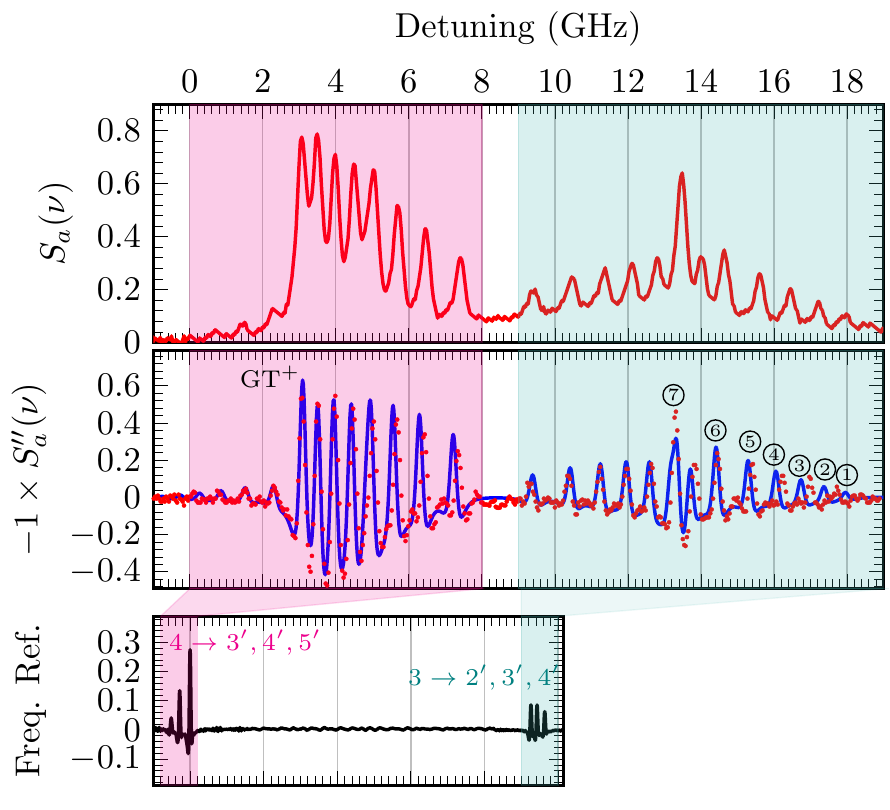}
\caption{Cs D$_2$ line spectrum of a $\sigma^+$ polarized radiation recorded at $B=2.40~$kG. Top panel: absorption spectrum; middle panel: SD of the absorption spectrum (red dots: experiment, blue solid line: theory); bottom panel: frequency reference.}
\label{fig:4}
\end{figure}

In figure~\ref{fig:4}, we show the absorption spectrum (top panel), and the SD spectrum (middle panel) recorded this time at $B=2.40$~kG from a NC illuminated with a $\sigma^+$ polarized laser radiation. The agreement between the experimental SD spectrum (middle panel, red dots) and the calculated one (blue solid line) is still remarkable.

In both Fig.~\ref{fig:3} and \ref{fig:4}, one can see that transitions $F_g=3\rightarrow F_e=2',3',4'$ (teal) and $F_g=4\rightarrow F_e=3',4',5'$ (magenta) split and shift due to the magnetic field. This later group of transition include the MI transitions $\textcircled{1}$ to $\textcircled{7}$ which are observable and well resolved. One can also observe that the amplitude of the MI transition  $\textcircled{7}$ is the strongest among fourteen transitions $F_g=3\rightarrow F_e=4,5$ (see also the inset of Fig.~\ref{fig:1}). The transition labelled GT$^+$ is the so-called ``guiding" transition for $\sigma^+$ transitions: its intensity remains the same in the whole range of magnetic fields from 0 to 10~kG and has a constant shift of 1.40~MHz/G \cite{sargsyanJETP2015,sargsyanJETPL2015}, while all other transitions experiences various changes.


\begin{figure}[h!]
\centering
\includegraphics[scale=1]{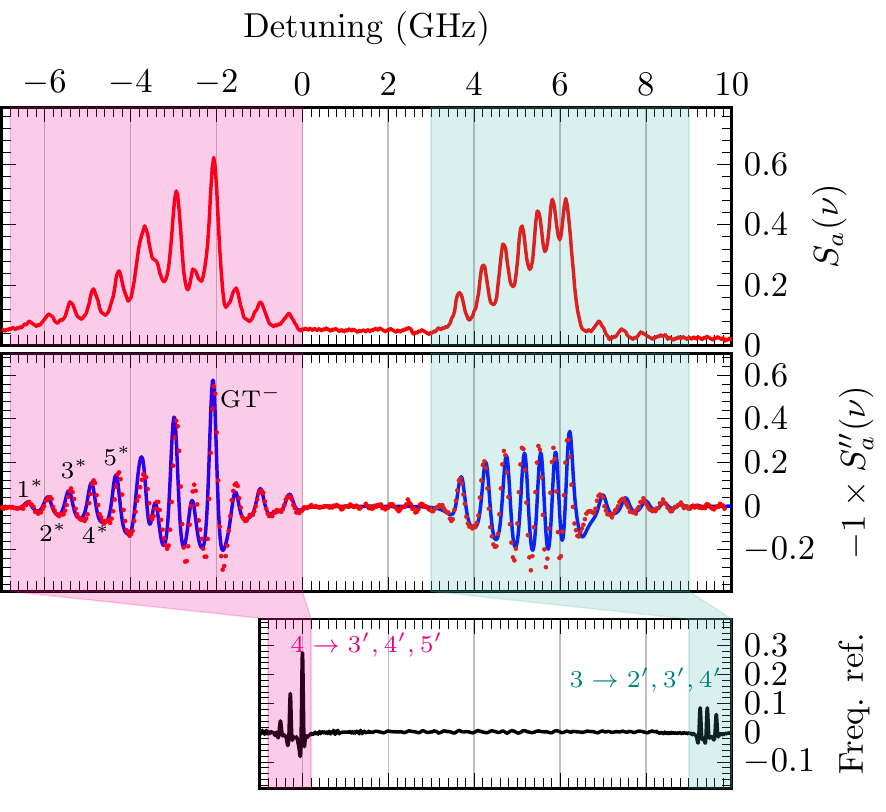}
\caption{Cs D$_2$ line spectrum of a $\sigma^-$ polarized radiation recorded at $B=1.50~$kG. Top panel: absorption spectrum; middle panel: SD of the absorption spectrum (red dots: experiment, blue solid line: theory); bottom panel: frequency reference.}
\label{fig:5}
\end{figure}

\subsection{MI transitions for a $\sigma^-$ excitation}

Figure~\ref{fig:5} shows the Cs D$_2$ line spectrum obtained with the same experimental parameters than that of Fig.~\ref{fig:3} with opposite circular polarization of the incident laser radiation ($\sigma^-$, $B=1.50~$kG). Again, the middle panel shows the SD spectrum obtained by processing the absorption spectrum (upper panel), and also shows a much improve spectral resolution. The lowest curve (black solid line) is the frequency reference spectrum obtained from another NC with $\ell\sim\lambda/2\approx 426$~ nm at zero magnetic field.


\begin{figure}[h!]
\centering
\includegraphics[scale=1]{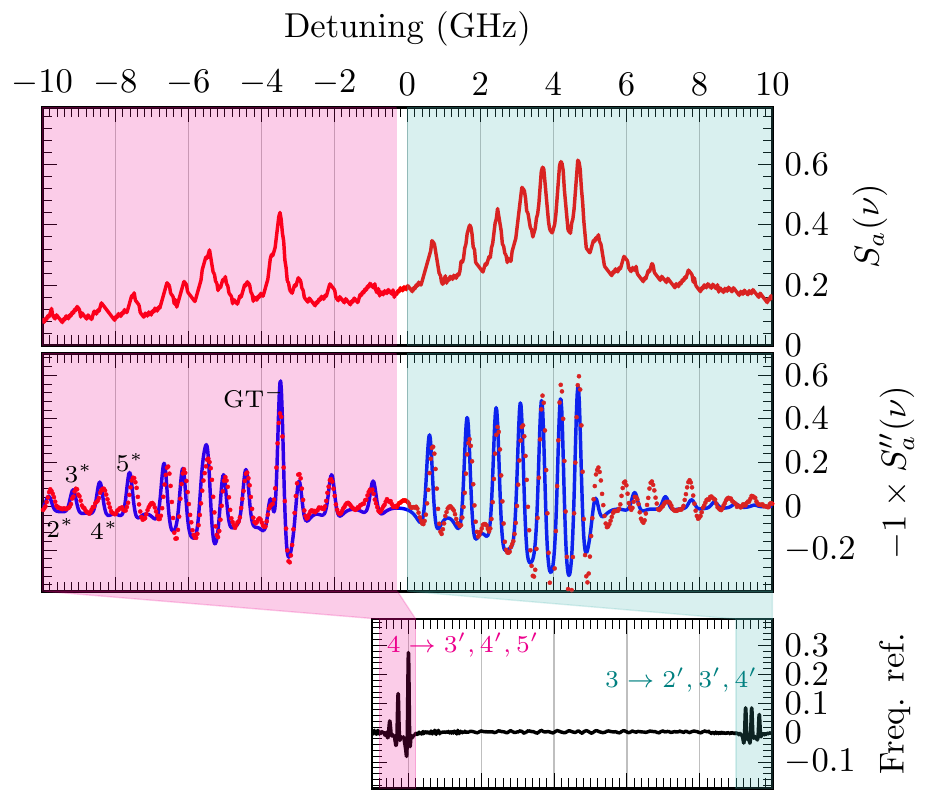}
\caption{Cs D$_2$ line spectrum of a $\sigma^-$ polarized radiation recorded at $B=2.40~$kG. Top panel: absorption spectrum; middle panel: SD of the absorption spectrum (red dots: experiment, blue solid line: theory); bottom panel: frequency reference.}
\label{fig:6}
\end{figure}

In figure~\ref{fig:6}, we show the absorption spectrum (top panel), and the SD spectrum (middle panel) recorded this time at $B=2.40$~kG from a NC illuminated with a $\sigma^-$ polarized laser radiation.
The agreement between the experimental SD spectrum (middle panel, red dots) and the calculated one (blue solid line) is also remarkable. In the SD spectra of figure~\ref{fig:5} and \ref{fig:6}, one can observe the five MI transitions labelled 1$^*$ to 5$^*$ as well as the guiding transitions GT$^-$ with a nice frequency resolution.

For magnetic fields $B>1.0$~kG, the MI transition $\textcircled{7}$, located on the high-frequency wing of the spectrum (see Figs.~\ref{fig:3} and \ref{fig:4}), and the MI transition 5$^*$, located on the low-frequency wing (see Figs.~\ref{fig:5} and \ref{fig:6}), are separated by more than 20~GHz, increasing with $B$. Therefore, to compare their intensities, we use the guiding transitions GT$^+$ and GT$^-$ which have the same intensities in the whole interval of magnetic fields of 0 to 10~kG: $I_{\text{GT}^+}= I_{\text{GT}^-}$. Since GT$^+$ is close to $\textcircled{7}$, while GT$^-$ is close to 5$^*$, then measuring the ratios $I_{\text{GT}^+}/I_{\textcircled{7}}$ and $I_{\text{GT}^-}/I_{5^*}$ yields $I_{\textcircled{7}}/ I_{5^*}=1.8\pm 0.1$ in both a magnetic field of 1.50~kG and 2.40~kG, which is consistent with the results of Fig.~\ref{fig:1} and Fig.~\ref{fig:7}.

\subsection{Circular dichroism}

\begin{figure}[h!]
\centering
\includegraphics[scale=1]{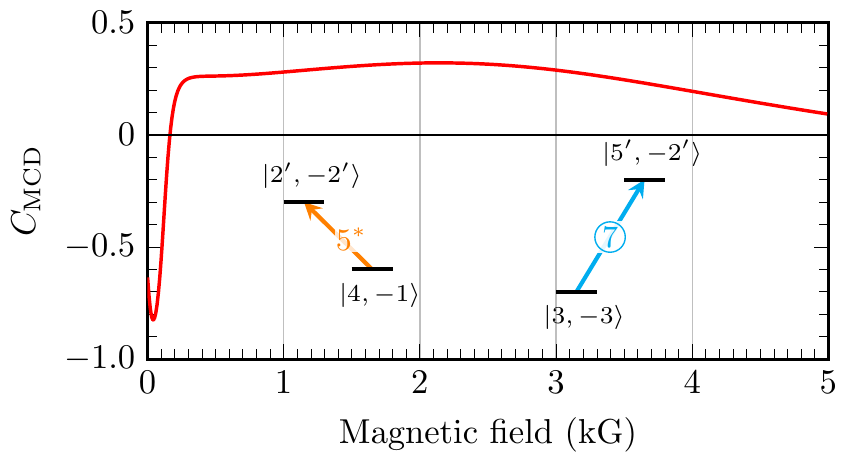}
\caption{Coefficient of magnetically induced circular dichroism for the strongest MI transitions for $\sigma^+$ polarized radiation and $\sigma^-$ polarized radiation, see eq.~\eqref{eq:CMICD}, as a function of the magnetic field. The diagrams for these two MI transitions are shown in inset. There is no circular dichroism at $B\approx 150~$G: $C_{\text{MCD}}=0$.}
\label{fig:7}
\end{figure}

To quantitatively describe MCD, we introduce the coefficient
\begin{equation}
C_{\text{MCD}} = \frac{I_{\textcircled{7}}-I_{5^*}}{I_{\textcircled{7}}+I_{5^*}},
\label{eq:CMICD}
\end{equation}
where $I_{\textcircled{7}}$ is the intensity of the transition $\textcircled{7}$ ($\sigma^+$ polarization), and $I_{5^*}$ is the intensity of the transition $5^*$ ($\sigma^-$), see Fig.~\ref{fig:7}. It is easy to see that the inequality $C_{\text{MCD}}  >0$, respectively $C_{\text{MCD}} < 0$, means that the large intensity of the transition line is reached in the case of $\sigma^+$ polarized radiation, respectively $\sigma^-$ polarized radiation. The equality $C_{\text{MCD}}=0$ means that transitions $\textcircled{7}$ and $5^*$ have the same intensity. As seen from Fig.~\ref{fig:7}, for $B>150$~G the intensity of the transition $\textcircled{7}$ is always lager than that of transition $5^*$. We refer to this effect as magnetically-induced circular dichroism of type-2.

\section{Conclusion}

    In this paper, we have described both the experimentally and theoretically another type of MCD, so called type-2 MCD, occuring on the Cs D$_2$ line. Comparison of the intensity of the strongest MI transitions (verifying $\Delta F=\pm2$) of the Cs D$_2$ line shows that the $\sigma^+$ MI transition $|3,-3\rangle\rightarrow|5',-2'\rangle$ is always greater than $|4,-1\rangle\rightarrow|2',-2'\rangle$ ($\sigma^-$ MI transition) in the range 0.15 to 5~kG, leading to another type of circular dichroism.  We refer to this unusual behavior as type-2 MCD. 
    
Magnetically induced circular dichroism of type-2 is also expected for other alkali atoms ($^{85}$Rb, $^{87}$Rb, $^{39}$K, etc.). For example, preliminary calculations have confirmed that the probability of the strongest $\sigma^+$ MI transition of $^{87}$Rb $|1,-1\rangle \rightarrow |3',0'\rangle$ is larger than the probability of the strongest $\sigma^-$ MI transition $|2,1\rangle \rightarrow |0', 0'\rangle$. For this reason, one could make use of MCD2 for effective applications of magneto-optical processes in atomic vapors \cite{budkerRMO2002}. Other preliminary calculations suggest that MICD1 and MICD2 could also be observed in the first fundamental series of $nS\rightarrow nP$ alkali D$_2$ lines, where $n = 3, 4, 5$ for Na, K, Rb, denotes the respective principle quantum number, which amounts to about 100 MI transitions. 
	
	Note that the recent development of a glass NC \cite{peyrotOL2019}, which is easier to manufacture than sapphire-made NC used in the present work, could make the nanocells and the above presented technique for study of unusual behavior of the MI transitions available for a wider range of researchers.
	
\section*{Acknowledgement}
The authors acknowledge A. Papoyan for fruitful discussions. This work was supported by the Science Committee of the Ministry of Education, Science, Culture and Sport of the Republic of Armenia (project no SCS 19YR-1C017).
\bibliography{PhysLettA2020.bib}

\end{document}